\documentclass[11pt]{article}
\usepackage{epsfig}
\title{\textbf{Ultraslow light propagation in an inhomogeneously
       broadened rare-earth ion-doped crystal}}
\author{E.~Baldit, K.~Bencheikh, P.~Monnier, J.~A.~Levenson and V.~Rouget
\\
\\
\emph{Laboratoire de Photonique et de Nanostructures LPN--CNRS UPR
020}\\
\emph{Route de Nozay, 91460 Marcoussis, France}\\
\emph{Email: kamel.bencheikh@lpn.cnrs.fr}}
\date{}

\begin{document}

\maketitle

\setlength{\baselineskip}{7mm}

\begin{abstract}
We show that Coherent Population Oscillations effect allows to
burn a narrow spectral hole (26\,Hz) within the homogeneous
absorption line of the optical transition of an Erbium ion-doped
crystal. The large dispersion of the index of refraction
associated with this hole permits to achieve a group velocity as
low as 2.7\,m/s with a transmission of 40\,\%. We especially
benefit from the inhomogeneous absorption broadening of the ions
to tune both the transmission coefficient, from 40\,\% to 90\,\%,
and the light group velocity from 2.7\,m/s to 100\,m/s.
\end{abstract}

\noindent Reduction of the optical group velocity by several
orders of magnitude relatively to the speed of light propagating
in vacuum was demonstrated in recent years in diverse media among
which Bose-Einstein condensate, cold atoms\,\cite{bec1,bec2},
vapors\,\cite{vapor1,vapor2} and
solids\,\cite{pr,ruby,alexandrite,qw}. In addition to the
strangeness of producing light propagating at speeds as low as few
m/s, Slow Light Propagation (SLP) is at the very heart of new
fundamental and applied fields of research in nonlinear and
quantum optics. From the nonlinear optical side, SLP allows to
strongly enhance the light-matter interaction time. Moreover, this
interaction time can be continuously tuned to produce optical
buffers and variable delay lines for optical networks.  From the
quantum optical point of view, SLP should allow, under specific
conditions, classical and quantum properties of an electromagnetic
field to be mapped into an atomic system\,\cite{polariton}. The
fundamental physical idea at the origin of SLP is the creation of
a very narrow spectral hole in the homogeneous absorption profile.
As stipulated by Kramers-Kr\"{o}nig relations, this narrow
spectral hole is accompanied by a strong dispersion of the index
of refraction inducing a low group velocity and an increase of the
transmission. These two aspects are crucial in the choice of the
atomic system and the coherent interaction inducing SLP.

The first direct demonstration of
SLP\,\cite{bec1,bec2,vapor1,vapor2,pr} was achieved via
Electromagnetically Induced Transparency (EIT)\,\cite{eit}. It was
originally implemented by applying a secondary control field to
eliminate the linear absorption of a resonant probe field through
an otherwise absorbing medium. The standard scheme for EIT is a
three-level $\Lambda$ system, where the probe field drives the
system from one of the ground states and the control field from
the second ground state. The width of the spectral hole burned in
the homogeneous absorption profile is proportional to the inverse
of the dephasing time of the ground state. Another physical effect
recently used to reduce the speed of light to as low as few tens
of meters per second is Coherent Population Oscillations
(CPO)\,\cite{ruby,alexandrite,qw}. In contrast to EIT, the
CPO\,\cite{boydbook,sargent} effect is easily achieved in a
two-level system which is excited by a "strong" pump field $E_p$
oscillating at $\omega_p$, along with a weak probe field $E_s$
oscillating at $\omega_s=\omega_p+2\pi\delta$. In this
configuration, the two-level system population oscillates at the
pump-probe beat frequency $\delta =(\omega_s-\omega_p)/2\pi$. Such
oscillations become significant when $\delta$ is less than
$1/T_{1}$, where $T_1$ is the relaxation lifetime of the excited
state. As a result a narrow spectral hole, whose width is
proportional to $1/T_1$, is induced in the homogeneous absorption
profile of the probe field. M. S. Bigelow \emph{et
al.}\,\cite{ruby} used this effect in a 7.25-cm-long ruby crystal
at room temperature to reduce the group velocity of light down to
57\,m/s with a spectral hole width of 36\,Hz and a rather low
total transmission of about 0.1\,\%. More recently, P.C. Ku
\emph{et al.}\,\cite{qw} showed in GaAs/AlGaAs semiconductor
quantum wells a high transmission of about 50\,\% for a group
velocity of 9600\,m/s. Though this value is low compared to the
speed of light in vacuum, it is still limited by the rather short
relaxation time $T_1 \sim 1$\,ns.

In this letter, we report on SLP induced by the CPO effect in a
crystal insulator doped with rare-earth ions. Thanks to the
long-lived excited state of the ions which is at the origin of the
very narrow spectral hole and of the steep dispersion of the
refraction index, we successfully achieved ultraslow light
propagation in a solid-state material. The corresponding group
velocities are comparable to those achieved in SLP via EIT in a
Bose-Einstein condensate\,\cite{bec1,bec2}. The optical
intensities we use are as weak as the intensities concerned in
Ref.\,\cite{bec1,bec2} and are orders of magnitude weaker than
those used in Ref.\,\cite{ruby,qw} for SLP via the CPO effect. In
contrast to Ref.\,\cite{ruby,alexandrite,qw}, we investigate in
details the influence of the power and the absorption coefficient
on the SLP. More importantly, we show that the inhomogeneous
broadening inherent to rare-earth ion-doped crystals, constitutes
an additional control parameter we are able to manipulate to tune
both the group velocity and the transmission coefficient.

\begin{figure}[tb]
\centerline{\includegraphics[angle=-90,width=\columnwidth]{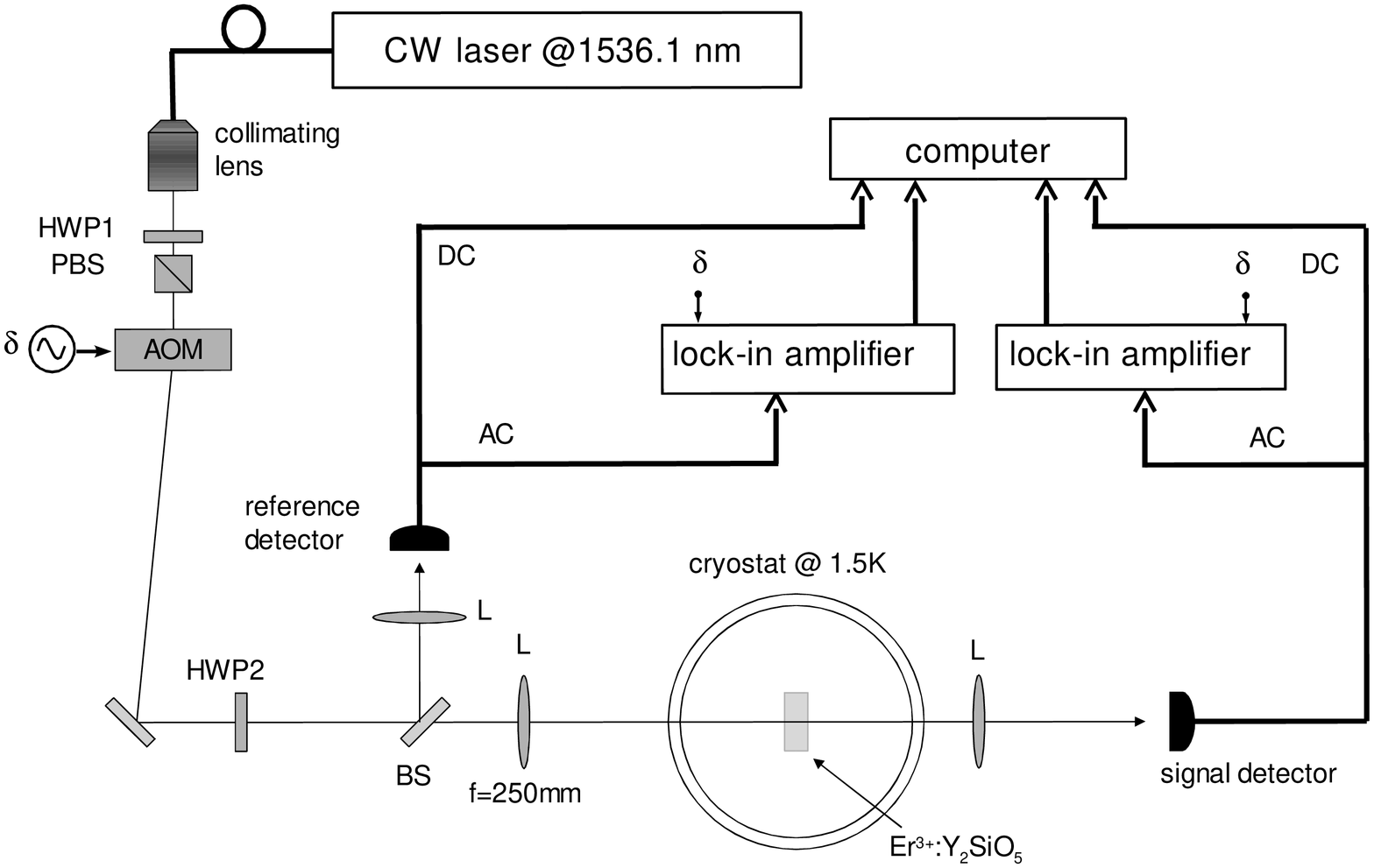}}
\caption{Scheme of the experimental setup:  HWP1, HWP2
half-wavelength plates; PBS polarizing beam splitter; AOM
acousto-optical modulator driven by a sinusoidal wave function
oscillating at $\delta$; BS beam splitter; L lens; AC Alternative
Current; DC Direct Current.} \label{setup}
\end{figure}

The crystal used in our experiments is supplied by Scientific
Materials Inc., and it consists in an Yttrium oxyorthosilicate
(Y$_{2}$SiO$_{5}$) lattice doped with 0.005\,at.\,\% of
$^{167}$Er$^{3+}$ ions. The optical transition between the
fundamental state $^{4}$I$_{15/2}$ and the excited state
$^{4}$I$_{13/2}$ of the Erbium ions is at 1536.1\,nm. The crystal
temperature is maintained at 1.5\,K in a liquid Helium cryostat in
order to reduce the homogeneous linewidth to the sub-MHz level.
However, due to the static electric field strength in the crystal
an inhomogeneous broadening of about
$\Gamma_{\mathrm{inh}}=1.3$~GHz is present and prevents from the
direct measurement of the homogeneous spectral line. Using the
spectral hole burning method, we measure the lifetime $T_{1}$ of
the excited state to be 8\,ms. Photon echo technique is used to
measure the dephasing time $T_{2}$ which is found to be about
3\,$\mathrm{\mu}$s at 1.5~K, a rather short value in comparison
with $T_{1}$.

From the theoretical point of view, the pump beam allows to
selectively excite a narrow portion of ions within the
inhomogeneous line: those in resonance with the pump frequency.
Such ensemble of ions can be described by the formalism developed
in Ref.\,\cite{ruby,boydbook}, where a two-level system is driven
by $E_{p}e^{-\imath \omega_{p}t}+E_{s}e^{-\imath \omega_{s}t}$
with $\omega_{s}=\omega_{p}\pm 2\pi \delta$. Following the
mathematical analysis in Ref.\,\cite{ruby,boydbook}, expressions
of the absorption $\alpha(\delta)$ and the index of refraction
$n(\delta)$ are derived for $T_2 \!<\!<\! T_1$ from the off
diagonal density matrix elements oscillating at $\delta$,
\begin{equation}\label{absorp}
\alpha(\delta)=\frac{\alpha_\circ}{1+S}
[1-\frac{S(1+S)}{(2\pi\delta T_1)^2+(1+S)^2}],
\end{equation}
and
\begin{equation}\label{indref}
n(\delta)=1+\frac{\alpha_\circ  c  T_1}{2 \omega_p} \frac{S}{1+S}
\frac{2\pi\delta}{(2\pi\delta T_1)^2+(1+S)^2}.
\end{equation}
The different parameters involved in Eqs.\,(\ref{absorp}) and
(\ref{indref}) are the saturation parameter $S$ defined as the
ratio of the pump intensity and the saturation intensity and the
unsaturated absorption coefficient $\alpha_\circ$ of the two-level
system. The Half Width at Half Maximum (HWHM) of the spectral hole
induced by the CPO effect is theoretically given by
\begin{equation}\label{hwhm}
\Gamma_{\mathrm{HWHM}}=\frac{1}{2\pi T_1}(1+S).
\end{equation}

The main concern of this letter is the determination of the group
velocity $v_g$ experienced by the probe field propagating in the
crystal of length $L$. It is deduced from the time delay $\tau$
accumulated by the probe, relatively to a reference beam. The
group velocity is given by $v_g=L/\tau$, where a theoretical
expression of the delay is easily obtained from
Eq.\,(\ref{indref}): {\setlength\arraycolsep{2pt}
\begin{eqnarray}\label{delay}
\tau(\delta)&\simeq& \frac{L}{c}
\frac{\omega_p}{2\pi}\frac{n(\delta)-n(-\delta)}{2\delta}\nonumber
\\
&=&\frac{\alpha_\circ T_1 L}{2} \frac{S}{1+S} \frac{1}{(2\pi\delta
T_1)^2+(1+S)^2},
\end{eqnarray}}
where we use the fact that $n \simeq n_0+(\omega_p/2\pi)\partial
n/\partial \delta$.

According to Eq.\,(\ref{hwhm}), the width of the spectral hole
associated with $T_{1}=8$\,ms is expected to be of the order of
few Hz. In this case, the frequency of the probe field is so close
to the pump frequency that the observation of the narrow spectral
hole is prevented by the relative frequency fluctuations of the
pump and the probe laser sources. However, it is possible to
overcome this problem by generating both the pump and the probe
fields from a single laser source whose intensity is modulated at
a frequency $\delta$ with a $1+m\cos(2 \pi \delta t)$ wave
function, where $m<1$ is the modulation depth. Indeed, the strong
mean value of the modulated laser signal plays the role of the
pump whereas the Fourier sidebands of the modulation at $\pm
\delta$ act as the weak probe. Such an approach has been
successfully used in Ref.\,\cite{ruby,alexandrite}.

The scheme of our experimental setup is depicted in
Fig.\,\ref{setup}. The light source we use is a 2-kHz linewidth,
continuous-wave fiber laser emitting at 1536.1\,nm. The
combination of the half-wave plate HWP1 and the polarizing
beamsplitter PBS allows to control the laser beam power. By
driving the rf power of an acousto-optic modulator (AOM) with a
$1+m\cos(2\pi\delta t)$ wave function, a 10\,\% intensity
modulation depth is applied to the first order diffracted field.
After the AOM, the beamsplitter BS allows to separate a reference
beam from the main beam which is sent through the crystal located
in the cryostat. Thanks to a 250-mm lens in front of the cryostat,
the main beam is focused down to a waist of 540\,$\mu$m in the
3-mm long crystal. The main and the reference beams are then
detected using InGaAs photodiodes. The AC components of the
photocurrents, proportional to the amplitude of the modulation,
are demodulated using lock-in amplifiers. They are then stored in
a computer, simultaneously with the DC components, after data
sampling and data averaging over more than 10000 samples.

\begin{figure}[h]
\centerline{\includegraphics[width=\columnwidth]{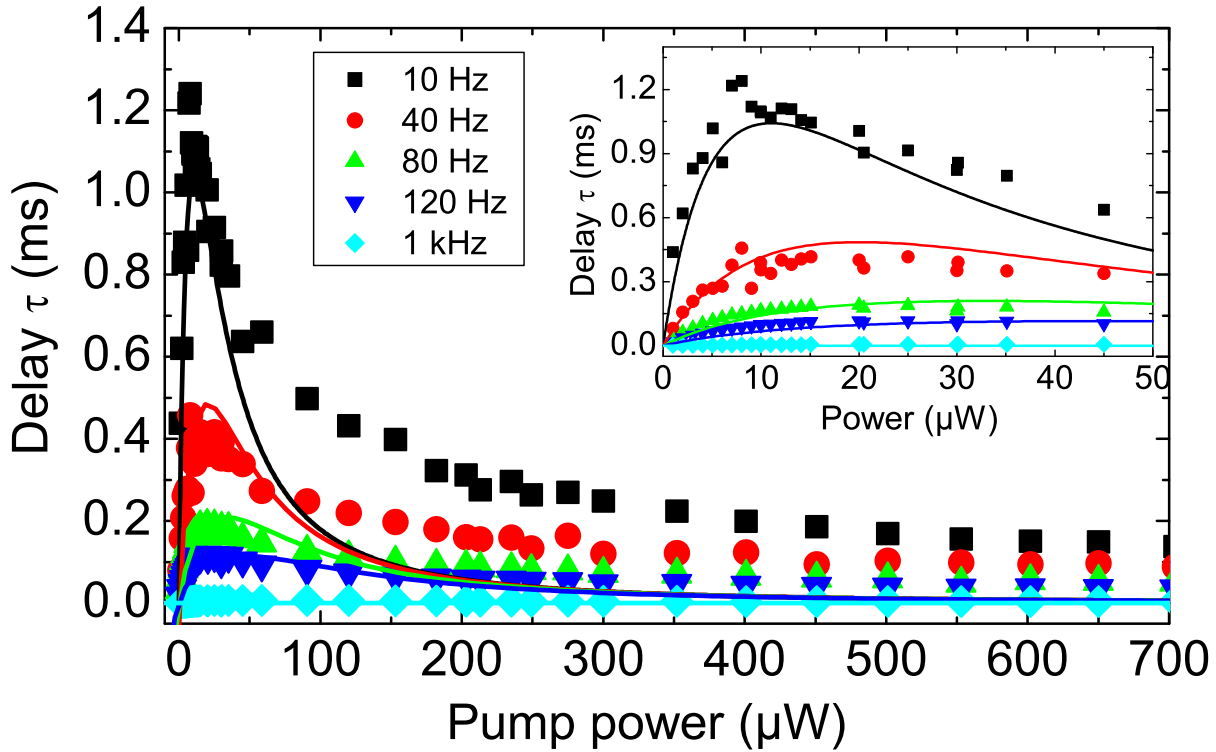}}
\caption{Experimental (points) and theoretical (continuous lines)
delay $\tau$ of the probe as a function of the pump power for
different modulation frequencies $\delta$. Inset: zoom at low pump
powers.} \label{delayvspower}
\end{figure}

Figure~\ref{delayvspower} shows the measured delay $\tau$ of the
weak sidebands for different modulation frequencies $\delta$ as a
function of the pump power. The pump field wavelength is tuned to
the maximum of the absorption in the inhomogeneous profile. The
continuous lines are the plots of the theoretical expression
Eq.\,(\ref{delay}) of the delay, where $T_1$=8\,ms, $L$=3\,mm. The
unsaturated absorption coefficient at the maximum of the
inhomogeneous line, $\alpha_\circ=6.5\,$cm$^{-1}$, is determined
by measuring the linear absorption in the crystal of a weak laser
beam as a function of its wavelength. Care is taken in tuning the
frequency of the laser over few GHz ($>\Gamma _{\mathrm{inh}}$)
within a sufficiently short time scale (5\,ms) in order to avoid
any saturation effect. As shown in Fig.\,\ref{delayvspower}, the
delay $\tau$ tends to zero in the low pump power limit,
$S\!<\!<\!1$, because the number of the excited ions is still not
enough to induce the coherent effect. When the pump power is
increased, $S\!>\!>\!1$, the delay $\tau$ decreases as well
because of the absorption saturation: the number of the excited
ions contributing to the CPO effect is reduced and the spectral
hole is power-broadened leading to a less sharp dispersion. The
discrepancy between theory and experiment when $S\!>\!>\!1$
\,\footnote{We experimentally checked that the increase of the
sidebands intensity as the pump power is increased, is not
responsible for the discrepancy in Fig.\,\ref{delayvspower}.} is
probably due to the inhomogeneous broadening\,\cite{sargent}.
Indeed, it modifies the saturation behavior of the Erbium ions
with respect to that of the homogeneously broadened two-level
system considered in the model. A maximum delay $\tau_{max}$ is
reached for an optimum value of $S(\delta)$ independent on
$\alpha_\circ$. It is $S\simeq$\,1/2 in the low frequency limit
($\delta < 1/T_1$) according to theory. Experimentally,
$\tau_{max}^{exp}=1.1\pm$\,0.1\,ms at $\delta=$~10~Hz which is in
a good agreement with the theoretical value
$\tau_{max}^{the}=$~1.04~ms derived from Eq.\,(\ref{delay}). The
delay $\tau_{max}^{exp}$ is obtained for a pump power of
12\,$\mu$W measured in front of the cryostat which corresponds to
an intensity of 2.1\,mW/cm$^2$ on the crystal. In the following,
we will work at this particular pump intensity in order to
maximize the CPO effect. This intensity is rather low in
comparison with former CPO experiments~\cite{ruby, alexandrite,qw}
where $\sim$\,kW/cm$^2$ are required, and is comparable to the
intensities involved in SLP in Bose-Einstein
condensate\,\cite{bec1,bec2}. The group velocity $v_g=L/\tau=$
associated with $\tau_{max}^{exp}$ is 2.7$\pm$0.2~m/s. To our
knowledge, it is the slowest group velocity ever achieved in a
solid. Indeed, according to our results, we reduce the group
velocity by a factor $n_g=c/v_g=10^8$. In comparison, the
reduction is less than $5\times10^6$ in the Ruby and Alexandrite
crystals \cite{ruby,alexandrite} and about $10^4$ in the
semiconductor quantum wells\,\cite{qw}.

\begin{figure}[h]
\centerline{\includegraphics[width=\columnwidth]{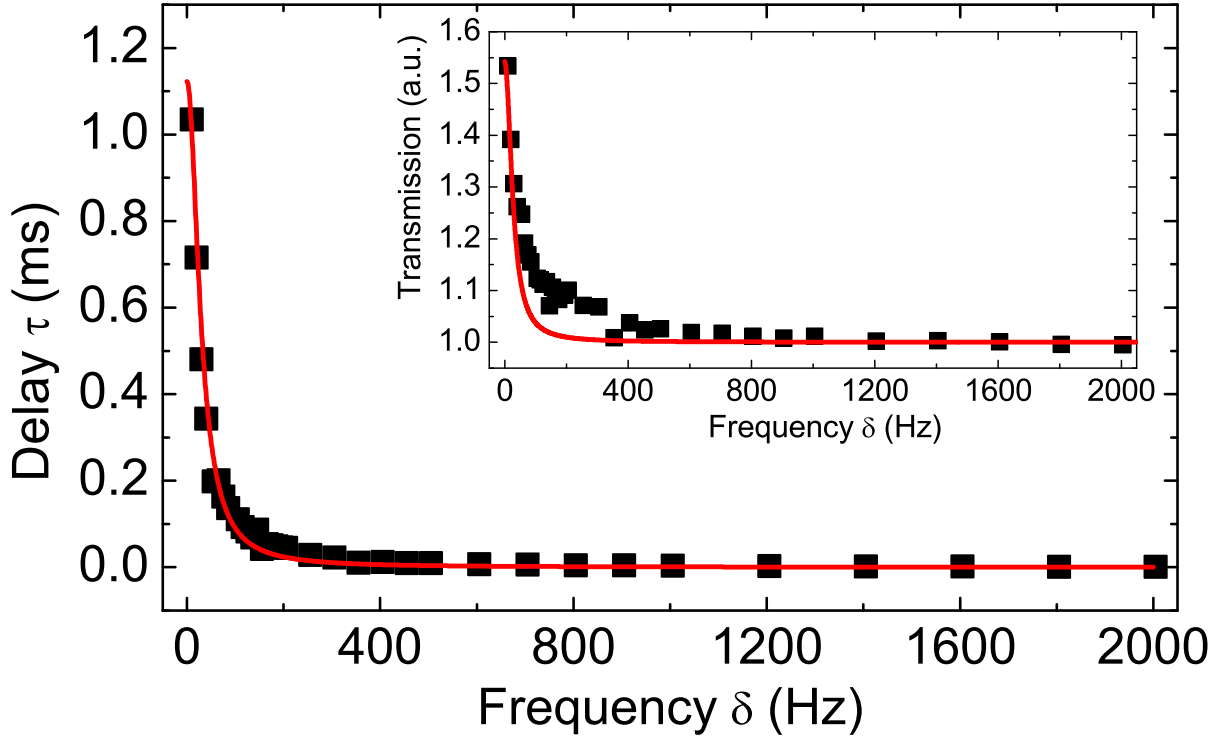}}
\caption{Experimental (squares) and theoretical (continuous lines)
delay $\tau$ of the probe  as a function of the modulation
frequency $\delta$ and for a fixed pump power (12\,$\mu$W) with
$S$=1/2. Inset: the corresponding normalized transmission.}
\label{delayvsfrequency}
\end{figure}

In order to determine the linewidth of the narrow spectral hole
which is at the origin of the large dispersion of the refraction
index and hence of the SLP, we measure the delay as well as the
transmission of the weak sidebands as a function of the modulation
frequency (the squares in Fig.\,\ref{delayvsfrequency}). Like in
the previous measurement, the wavelength of the pump is tuned to
the maximum of the absorption. The inset in
Fig.\,\ref{delayvsfrequency} is the transmission of the weak
sidebands normalized to the transmission that is observed in a
hole burning experiment. Indeed, a spectral hole is burned by the
pump field in the inhomogeneous transmission profile of the probe
field. Due to CPO effect, an additional increase of the
transmission is observed when $\delta$ is smaller than $1/T_1$.
This increase is about 55\,\% at $\delta$=~10~Hz. At frequencies
$\delta$ larger than $1/T_1$, the effect of CPO becomes
negligible. The continuous lines in Fig.\,\ref{delayvsfrequency}
are the theoretical plots derived from Eq.\,(\ref{absorp}) and
Eq.\,(\ref{delay}) without adjustable parameters. We can deduce
from the inset in Fig.\,\ref{delayvsfrequency} a 26-Hz HWHM of the
spectral hole, which is rather close to the theoretical value of
30\,Hz.

\begin{figure}[h]
\centerline{\includegraphics[width=\columnwidth]{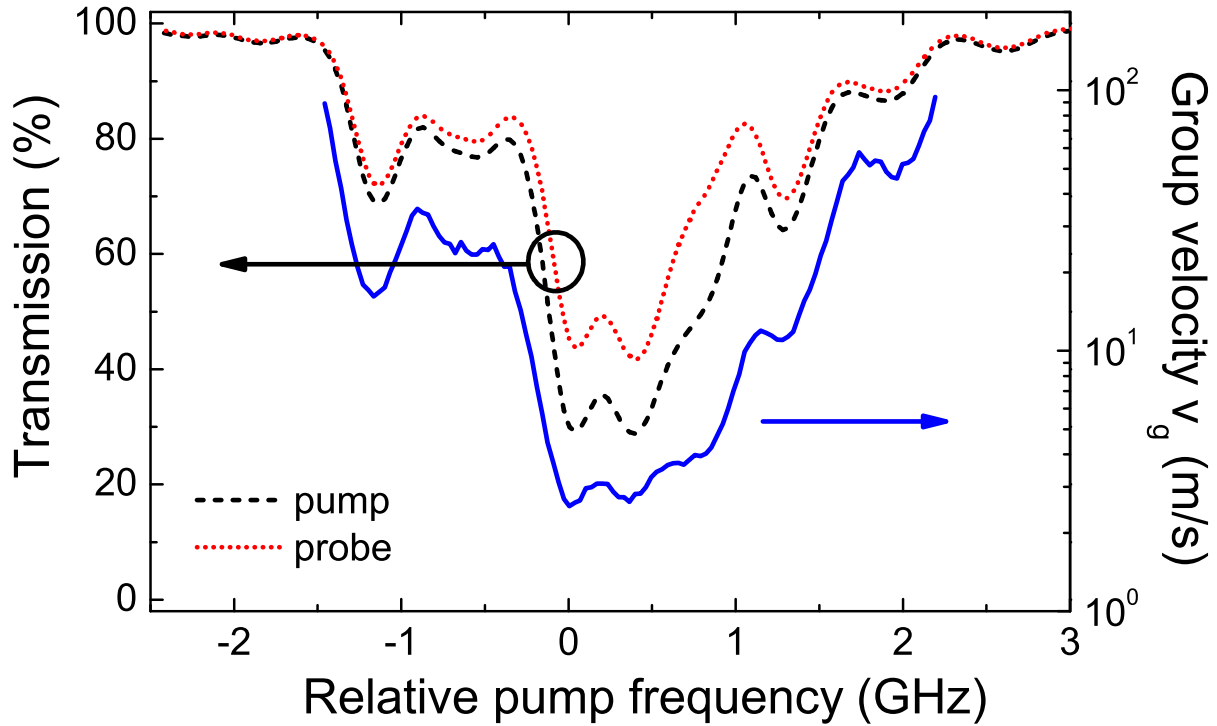}}
\caption{The experimental group velocity (continuous line), pump
(dashed line) and probe (dotted line) transmissions as a function
of the pump frequency. The modulation frequency $\delta$ is fixed
at 10\,Hz and the pump power is such that $S=$1/2.}
\label{delayvsfpump}
\end{figure}

One of the fundamental point we are addressing in this letter is
the experimental exploration and exploitation of the dependance of
the group velocity reduction on the absorption coefficient.
Indeed, unlike the other materials that have been investigated up
to now \cite{ruby,alexandrite,qw}, our crystal shows an
inhomogeneous absorption broadening. By tuning the pump frequency
over this inhomogeneous profile, we are able to address different
Erbium ions concentrations and hence different absorption
coefficients. In Fig.\,\ref{delayvsfpump}, the continuous line
represents the measured light group velocity as a function of the
pump frequency detuning relative to the pump wavelength at the
maximum of the absorption. We also plot the transmission of the
pump field (dashed line) as well as the transmission of the probe
field (dotted line). The frequency of the probe is detuned by
$\delta=10$\,Hz from the pump frequency. The shape followed by the
group velocity is similar to that of the transmission, clearly
indicating the dependance of $v_g$ on the absorption strength in
the Erbium-doped crystal. However, a careful analysis of the
results shows that $v_g$ is not proportional to the inverse of the
absorption. This might be due to the inhomogeneous broadening.
From the practical point of view, Fig.\,\ref{delayvsfpump}
indicates that the inhomogeneous broadening is an additional
parameter which allows the engineering of the group velocity.
Indeed, thanks to the inhomogeneous broadening in our crystal,
$v_g$ can easily be varied from 3\,m/s to 100\,m/s by tuning the
pump wavelength. The highest SLP is achieved at the maximum of the
absorption with a probe transmission of 40\,\%. Moreover, an
important reduction of the group velocity is still obtained for
high probe transmission. For example, when the relative pump
frequency is set at 2\,GHz (see Fig.\,\ref{delayvsfpump}), a group
velocity of 58\,m/s is measured with a transmission of about
88\,\%.

It is worth noting that the delay $\tau$ is proportional to the
product $\alpha_\circ T_1 L$. Consequently, the reduction of the
group velocity  $n_g$ varies like the product $\alpha_\circ T_1
c$. The slowest light propagation will be then obtained in
long-lived atomic systems with high absorption coefficient.
However, in order to obtain a non negligible transmission, $T\sim
e^{-\alpha_\circ L}$, one has to compensate for the large
coefficient $\alpha_\circ$ by choosing a short medium. It is for
example possible to obtain a 88\,\% transmission in the
Erbium-doped crystal with the lowest group velocity, 3\,m/s,  by
using a crystal having a thickness of about 200\,$\mu$m pumped at
the maximum of the absorption profile \footnote{We checked this
dependance on $L$ by comparing experimental results obtained in a
1- and a 3-mm long crystals.}. This means that transparency can be
reached while keeping the group velocity at its minimum value.

In conclusion, the implementation of CPO in Erbium-doped crystals
enables to achieve ultra SLP. The pump intensities involved are
six order of magnitude smaller than in former CPO experiments
\cite{ruby, alexandrite,qw} and comparable to the one involved in
EIT in ultracold atomic gas\,\cite{bec1,bec2}. We demonstrate that
the inhomogeneous broadening, an inherent effect of the
crystalline matrix, does not degrade the slow light propagation
effect. It can instead be used to spectrally tune both the slow
light propagation effect and the transmission. Group velocities
ranging from 3~m/s to 100~m/s are achieved with transmission
efficiencies ranging from 40\% to 90\%.

We are grateful to V. Crozatier, I. Lorger\'{e}, F. Bretenaker,
J.-L. Le Gou\"{e}t, O. Guillot-No\"{e}l and P. Goldner for their
collaboration in the different spectroscopic measurements. We also
thank R. Raj, N. Belabas and A. Giacomotti for helpful comments on
the manuscript.

\end{document}